\documentclass[preprint,12pt]{elsarticle}
\usepackage[T1]{fontenc}

\usepackage{amsmath,amssymb,amsfonts}
\usepackage{booktabs}
\usepackage{tikz}
\usetikzlibrary{shapes.geometric, positioning, arrows.meta, calc}
\usepackage[utf8]{inputenc}
\usepackage{graphicx}
\usepackage{subcaption}
\usepackage{url}
\usepackage{hyperref}
\usepackage{balance}
\usepackage{placeins}
\begin{document}

\begin{frontmatter}

\title{Model-Driven Requirements Configuration with Three-Valued Uncertainty Scoring}

\author[1]{Ahmed Ibrahim\corref{cor1}}
\ead{aibrah64@uwo.ca}
\cortext[cor1]{Corresponding author}

\address[1]{Western University, London, Canada}

\begin{abstract}
\textbf{Context:} Large Language Models (LLMs) offer natural-language flexibility for automated requirements elicitation but frequently generate structurally invalid requirements and logical inconsistencies, lacking formal correctness guarantees.

\textbf{Objectives:} This study aims to eliminate logical inconsistencies and enforce structural conformance in LLM-generated requirements while quantifying the LLM's pre-validation decision uncertainty within a formal domain model.

\textbf{Methods:} We present a neuro-symbolic multi-agent architecture that operationalizes the Object-Oriented Method for Requirements Authoring and Management (OOMRAM) lattice. The LLM acts as a non-deterministic heuristic for lattice traversal, while a deterministic symbolic validator enforces all structural constraints. We introduce a three-valued $\langle T, I, F \rangle$ (Truth, Indeterminacy, Falsity) framework to classify and score the LLM's requirement decisions before and after validation.

\textbf{Results:} Evaluated across 37 natural-language project visions in eleven application families, the system completely eliminated structural inconsistencies in 35 out of 37 cases (94.6\%), with the remaining two containing only 6 unresolved structural errors (0.39\% of decisions) due to iteration limits. Three-valued analysis revealed that 24.7\% of all decisions are indeterminate (structurally valid but discretionary choices not explicitly mandated by the stakeholder).

\textbf{Conclusion:} Offloading structural integrity to a deterministic symbolic layer successfully guarantees structural conformance, while the three-valued classification provides a formal way to measure neural uncertainty, facilitating safe LLM deployment in formal requirements engineering.
\end{abstract} 


\end{frontmatter}

\section{Introduction}
\label{sec:introduction}

Requirements engineering (RE) is costly and error-prone. Reusing validated requirements from software product lines reduces effort and improves quality~\cite{mannion2000representing, clements2001software}. In requirements modelling, formal domain formalisms, such as the Object-Oriented Method for Requirements Authoring and Management (OOMRAM)~\cite{ibrahim2004, ibrahim2005thesis}, represent an application family as a strict mathematical structure. For instance, OOMRAM organizes requirements into a formal lattice where variability is expressed through discriminants. However, traditional formal methods typically demand exact requirement identifiers and rigid syntax, creating an "exact-match bottleneck" that prevents analysts from exploring options using natural-language project descriptions.

Recent literature reviews and empirical studies demonstrate that Large Language Models (LLMs) are increasingly applied to RE tasks, including automated requirement generation, tracing, and elicitation from natural language text~\cite{rosa2025largellmsre, lang2026generative, zadenoori2025largellms, Puchleitner2025, Masoudifard2024Leveraging}. In this paper, we refer to such a free-text stakeholder description as a project vision (or simply vision). A vision can be anything from a few keywords to several paragraphs, as long as it spells out the desired functionality and constraints in ordinary language. A short smart‑home illustration would be: "A voice‑controlled home automation system that adjusts lighting and room temperature based on occupancy and time of day, with a backup battery for power outages." However, pure LLM approaches are prone to logical inconsistencies: they invent non-existent identifiers, ignore mandatory constraints, or combine mutually exclusive options~\cite{khan2024llm4re, ji2023survey, Zhang2024LLM}. Most NLP4RE approaches lack formal correctness guarantees and are validated only in laboratory settings~\cite{Norheim2024}.

To address this gap, this paper proposes a novel neuro-symbolic multi-agent architecture that directly operationalizes a formal domain model, using the OOMRAM lattice~\cite{ibrahim2004, ibrahim2005thesis} as our primary exemplar. Within this framework, the LLM acts as a non-deterministic heuristic for lattice traversal, interpreting the free-text vision, while a separate, deterministic symbolic validator enforces all structural constraints. This coupling ensures that logically inconsistent requirement combinations are eliminated because they cannot bypass the symbolic verification layer. While the framework cannot prevent an LLM from selecting a semantically odd requirement, it guarantees that any selected specification is a logically valid instance of the domain metamodel.

\textbf{Scope clarification.} It is important to note that the system targets the \emph{feature-selection phase} of product-line requirements elicitation~\cite{clements2001software}. Given a pre-engineered OOMRAM lattice for an application family (e.g., eRecord Keeping or Smart Home), the goal is to identify which requirements from that lattice are mandated or compatible with a stakeholder's natural-language vision—a structured configuration task, not generative requirements authoring from scratch. This scoping decision enables formal structural guarantees that are not achievable in open-ended generation.

However, this study argues that simply labelling selections as binary "correct" or "incorrect" fails to capture a critical engineering distinction. While some requirements are strictly mandated by the project vision, others are structurally valid but discretionary, representing free choices among equally acceptable alternatives, such as selecting a specific communication protocol from a set of available adapters.

The main contributions of this paper are fourfold. First, we present a neuro-symbolic architecture that makes OOMRAM practical for natural-language elicitation by offloading structural logic to a symbolic layer. Second, we introduce a three-valued evaluation framework ($\langle T, I, F \rangle$) that quantifies the difference between mandated requirements, valid but indeterminate selections, and structural falsities. Third, we conduct an empirical evaluation on 37 project visions across eleven families, demonstrating that the system reduces structural falsity to 0.39\% under a lightweight model and achieves 100\% convergence and 0\% structural falsity under a frontier model. Fourth, we provide a comparative analysis with standard LLM baselines and a formal proof of validator convergence.

The remainder of this paper is organized as follows. Section~\ref{sec:related} reviews related work in NLP4RE and neuro-symbolic AI. Section~\ref{sec:methodology} defines the OOMRAM lattice and the multi-agent architecture. Section~\ref{sec:evaluation} presents the empirical results, including the three-valued uncertainty analysis and scalability study. Section~\ref{sec:discussion} discusses implications for practice and threats to validity. Section~\ref{sec:conclusion} concludes.

% ==================== 
% Research Questions
% ==================== 
\subsection{Research Questions}
\label{sec:rqs}

This paper addresses the following research questions:
\begin{enumerate}
    \item \textbf{RQ1 (Structural Conformance):} To what extent can a symbolic validator, integrated into an LLM-driven requirements elicitation loop, eliminate logical inconsistencies and enforce structural conformance in the final specification?
    \item \textbf{RQ2 (Three-Valued uncertainty):} How many of the LLM's decisions are truly indeterminate, that is, structurally correct but not something the vision explicitly demands, and does this change much from one application family to another?
    \item \textbf{RQ3 (Baseline comparison):} Compared to an unconstrained LLM (without validator), how many structural violations does our system prevent?
    \item \textbf{RQ4 (Scalability):} Does the traversal time scale linearly with the number of discriminant nodes, as claimed by the frontier-based navigation strategy?
\end{enumerate}
Each research question is answered in Section~\ref{sec:evaluation}: RQ1 via the final falsity count (Table~\ref{tab:tif_scores}); RQ2 via the indeterminacy proportion and family breakdown; RQ3 via the ablation analysis; RQ4 via scalability measurements.

\section{Related Work}
\label{sec:related}

Our work draws on product‑line requirements engineering, neuro‑symbolic AI, multi‑agent systems, formal verification, and three-valued logic.

\subsection{OOMRAM and Feature Modeling}
\label{sec:oomram_fm}

Feature Models (FMs) and the Feature-Oriented Domain Analysis (FODA) framework proposed by Kang et al.~\cite{kang1990feature} serve as the established standard for variability modeling in software product lines. Traditional feature models structure software options into a hierarchy or tree with mandatory, optional, alternative (XOR), and OR features.

Building upon foundations in requirement reuse, the Object-Oriented Method for Requirements Authoring and Management (OOMRAM) was introduced by Ibrahim~\cite{ibrahim2004, ibrahim2005thesis} to model reusable requirements in application families. OOMRAM was chosen specifically for this work because its \textbf{object-oriented lattice structure} (formalized as a Directed Acyclic Graph) handles multiple inheritance and cross-cutting requirement constraints more naturally than standard strictly tree-based Feature Models. In an OOMRAM lattice, a single requirement node can inherit from multiple parent discriminants, capturing shared software capabilities across distinct subsystem branches. The structural rules governing single-adaptors (SA), multiple-adaptors (MA), core nodes, and options in OOMRAM serve as the \textbf{standard constraints for this family of lattices}, providing a rich graph benchmark for formal constraint verification.

\subsection{LLMs and Neuro-Symbolic Systems in RE}
\label{sec:llm_re}

While formalisms like OOMRAM~\cite{ibrahim2004, ibrahim2005thesis} offer a rigorous lattice‑based structure for managing variability, their original retrieval mechanisms demand exact requirement identifiers. This "exact‑match bottleneck" prevents practical adoption in natural‑language settings.
Several recent efforts have brought LLMs to bear on core RE tasks (elicitation, classification, and traceability)~\cite{Alturayeif2026, rosa2025largellmsre, lang2026generative, zadenoori2025largellms, Puchleitner2025, Masoudifard2024Leveraging}.
However, the same studies consistently report logical inconsistencies, hallucinations, and an absence of formal guarantees~\cite{khan2024llm4re, ji2023survey}.
Techniques such as retrieval‑augmented generation (RAG)~\cite{Ayala2024}, iterative grounding~\cite{Eghbali2024De-Hallucinator}, and layered mitigation frameworks~\cite{Hiriyanna2025Multi-Layered} can suppress some errors, but their protection remains heuristic. They stop short of the deterministic enforcement that high‑assurance RE demands.

Neuro‑symbolic approaches pair neural flexibility with symbolic correctness~\cite{chaudhuri2021neurosymbolic, garcez2020}. Notable examples include LLM‑guided controller synthesis~\cite{Bayat2025}, where temporal logic constraints guide neural search, and formal verification of LLM‑produced code~\cite{Councilman2025, Yang2025b}, which translates LLM outputs into intermediate representations for constraint solving (e.g., using Z3). In particular, HaVen~\cite{Yang2025b} and similar approaches~\cite{Wang2024} verify logic ex-post, checking generated specifications against separate safety rules. In contrast, our approach embeds the symbolic constraints of the domain lattice directly into the generative loop, structurally prohibiting invalid selections from advancing beyond the navigator phase.
Multi‑agent RE frameworks have been built for stakeholder simulation~\cite{Ataei2024}, debate‑based refinement~\cite{oriol2025multiagent}, and collaborative privacy threat modelling~\cite{bissoli2026benchmarking}. Critically, none of them incorporate a deterministic constraint checker.
Formal methods for business‑constraint verification~\cite{Stoica2024Formal} and static model checking~\cite{Somogyi2023Verifying} deliver rigorous guarantees, but they operate outside LLM pipelines.

Three-valued logic frameworks add an explicit indeterminacy component (I) alongside truth (T) and falsity (F). Earlier three‑valued logics laid the groundwork for reasoning under partial information~\cite{Blikle1988, Konikowska1988}, and three-valued classification has since been applied to trust verification in ad‑hoc networks~\cite{Baharloo2023} and medical decision support~\cite{Khushal2025}.

Our work embeds a deterministic constraint checker directly in an LLM agent loop and evaluates each proposal with $\langle T,I,F \rangle$ scoring, coupling formal logical inconsistency elimination with fine‑grained uncertainty quantification.

% ================== 
% Comparative Table
% ================== 
\subsection{Comparative Analysis}
\label{sec:comparison}

Table~\ref{tab:comparison} summarizes how our system differs from prior work across five dimensions: formal guarantees, LLM usage, uncertainty quantification, multi-agent orchestration, and empirical validation scale.

\begin{table*}[htbp]
\centering
\caption{Comparison with related approaches.}
\label{tab:comparison}
\setlength{\tabcolsep}{3pt}  % Professional tight spacing
\begin{tabular}{@{}lccccc@{}}
\toprule
\textbf{Approach} & \shortstack{\textbf{Formal}\\\textbf{Guarantee}} & \textbf{LLM} & \shortstack{\textbf{Uncertainty}\\\textbf{Quant.}} & \shortstack{\textbf{Multi-}\\\textbf{Agent}} & \textbf{Visions} \\
\midrule
Norheim et al.~\cite{Norheim2024} & No & Yes & No & No & 5 \\
Vogelsang~\cite{Vogelsang2024} & No & Yes & No & No & -- \\
Wang et al.~\cite{Wang2024} & No & Yes & No & No & 20 \\
Bayat et al.~\cite{Bayat2025} & Yes (ctrl) & Yes & No & Yes & 3 \\
Councilman et al.~\cite{Councilman2025} & Yes (code) & Yes & No & No & 28 \\
Yang et al.~\cite{Yang2025a} & Yes (spec) & Yes & No & No & 2 \\
HaVen~\cite{Yang2025b} & Yes (HDL) & Yes & No & No & -- \\
Baharloo et al.~\cite{Baharloo2023} & No & No & Yes (neutr.) & No & -- \\
Khushal~\cite{Khushal2025} & No & No & Yes (3-val.) & No & -- \\
\textbf{This work} & \textbf{Yes (lattice)} & \textbf{Yes} & \textbf{Yes (T,I,F)} & \textbf{Yes} & \textbf{37} \\
\bottomrule
\end{tabular}
\end{table*}

\section{Neuro-Symbolic Multi-Agent System}
\label{sec:methodology}

\subsection{Overview of OOMRAM}
\label{sec:oomram}

The Object‑Oriented Method for Requirements Authoring and Management (OOMRAM) was originally designed to support systematic reuse of requirements across families of related software applications~\cite{ibrahim2004, ibrahim2005thesis}. It extends Mannion's earlier MRAM method~\cite{Mannion1999} and rests on three main concepts: application families, requirement lattices, and discriminants.

\textbf{Application families}. An \emph{application family} is a set of systems that share a common domain and a common core of functionality, while differing at well‑defined points of variability. Typical examples include record‑keeping systems, smart home controllers, and automotive infotainment systems.

\textbf{Requirement lattice}. All requirements of a family are organized as a directed acyclic graph (a lattice). Each node represents a reusable requirement object that may have several parents (multiple inheritance) and several children. The lattice captures both mandatory (core) and optional relationships.

\textbf{Discriminants.} Variability is expressed through three types of \emph{discriminant} nodes that behave conceptually like UI selection widgets to enforce constraints.  First, a \emph{single‑adaptor (SA)} node requires that exactly one child must be selected from a set of mutually exclusive alternatives, behaving like a radio button group (e.g., choosing either an Oracle or PostgreSQL database backend, but not both).  Second, a \emph{multiple‑adaptor (MA)} node requires that at least one child (and potentially several) must be selected, acting like a mandatory checkbox group (e.g., selecting supported export formats: PDF, CSV, or both).  Third, an \emph{option} node allows a child to be selected or omitted independently, conceptually behaving like an optional toggle switch.

Core nodes, by contrast, are strictly mandatory whenever their parent is selected. The lattice structure is used instead of a simple tree to capture complex, cross-cutting dependencies, such as a single requirement being a child of multiple different options (multiple inheritance).

\begin{figure}[htbp]
\centering
\begin{tikzpicture}[
    % Ultra-compact settings for column width
    node distance=0.8cm and 0.05cm, 
    every node/.style={
        draw, 
        rounded corners=1pt, 
        thick, 
        align=center, 
        font=\scriptsize, % Smallest readable font for columns
        inner sep=2pt,    % Reduces padding inside boxes
        minimum height=0.6cm,
        minimum width=1.5cm % Narrow boxes to save horizontal space
    },
    core/.style={fill=blue!10},
    sa/.style={fill=green!10, minimum width=2.4cm},
    ma/.style={fill=orange!5, minimum width=2.4cm},
    option/.style={dashed, fill=gray!2, minimum width=1.8cm},
    % Compact arrows
    >={Stealth[length=1.2mm]},
    thin
]

% --- NODES ---

% Root - Top Level
\node[core] (root) {1: Core \\ Record keeping};

% Level 2 - Parent branches (positioned closer to center)
\node[sa, below left=0.8cm and 0.4cm of root] (sa) {1.4 SA \\ Admin channel};
\node[ma, below right=0.8cm and 0.4cm of root] (ma) {3.3 MA \\ Link to info};

% Level 3 - The Row (Ordered to fit between branches)
% Position the central "Option" node first
\node[option, below=2.2cm of root] (opt) {3.3.3 Option \\ Other docs};

% Children of 1.4 (Left side)
\node[left=0.1cm of opt] (fax) {1.4.2 \\ Fax};
\node[left=0.1cm of fax] (email) {1.4.1 \\ Email};

% Children of 3.3 (Right side)
\node[right=0.1cm of opt] (notes) {3.3.1 \\ Notes};
\node[right=0.1cm of notes] (emails) {3.3.2 \\ Emails};

% Tiny text below
\node[draw=none, below=0.3cm of opt, minimum height=0pt, font=\tiny] (child) {optional child};

% --- EDGES ---

\begin{scope}[->]
    \draw (root) -- (sa);
    \draw (root) -- (ma);

    % Connections for SA
    \draw (sa) -- (email);
    \draw (sa) -- (fax);

    % Connections for MA
    \draw (ma) -- (opt); % Short, non-crossing path
    \draw (ma) -- (notes);
    \draw (ma) -- (emails);

    % Dashed child
    \draw[dashed] (opt.south) -- (child.north);
\end{scope}

\end{tikzpicture}
\caption{A fragment of an OOMRAM lattice showing core (blue), single‑adaptor (SA, green), multiple‑adaptor (MA, orange), and optional (gray, dashed) nodes. Parent‑child relationships are indicated by arrows.}
\label{fig:oomram-lattice}
\end{figure}
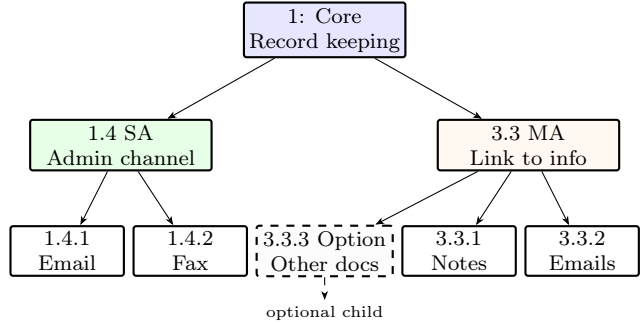

Figure~\ref{fig:oomram-lattice} shows a small lattice fragment. Core nodes (e.g., requirement 1) are mandatory when selected; SA discriminants (e.g., 1.4) require exactly one child; MA discriminants (e.g., 3.3) require at least one child; optional nodes (e.g., 3.3.3) may be included or omitted.

\textbf{Original retrieval bottleneck}. In the original OOMRAM tool, an analyst navigated the lattice by manually choosing discriminants or by providing exact requirement identifiers.  Natural‑language project visions were not supported.  As a result, the framework's reuse potential remained largely untapped, a limitation that our neuro‑symbolic architecture directly addresses.

For a complete description of OOMRAM, including its object‑oriented requirement representation, versioning, and repository design, we refer the reader to~\cite{ibrahim2004, ibrahim2005thesis}. The next subsection formalizes the lattice model that forms the backbone of our symbolic validator.

\subsection{Lattice Model}
\label{sec:lattice}

We formally define an OOMRAM requirement lattice as a 4-tuple:
\begin{equation}
\mathcal{L} = (V, E, \mathcal{T}, \Phi)
\label{eq:lattice_tuple}
\end{equation}
where $V$ is the finite set of requirement nodes, $E \subseteq V \times V$ is the set of directed parent-child edges forming a Directed Acyclic Graph (DAG), $\mathcal{T}: V \to \{\text{Core}, \text{SA}, \text{MA}, \text{Option}\}$ assigns a structural node type to each node, and $\Phi$ is the set of global lattice invariants.

Let $\mathcal{S} \subseteq V$ be the set of requirements currently selected by the agent during traversal. For any node $v \in V$, we define its parent set and children set respectively as:
\begin{align}
\mathrm{parent}(v) &= \{p \in V \mid (p, v) \in E\} \\
\mathrm{children}(v) &= \{c \in V \mid (v, c) \in E\}
\end{align}

A selection set $\mathcal{S}$ is structurally valid if and only if it satisfies the global consistency predicate $\Psi(\mathcal{S})$, defined by the conjunction of four axiomatic structural invariants:

\noindent\textbf{1. Core Invariant (Mandatory Selection):} Any Core requirement node must be selected whenever at least one of its parent nodes is active in $\mathcal{S}$:
\begin{equation}
\forall v \in V,\; \bigl(\mathcal{T}(v) = \text{Core} \land \mathrm{parent}(v) \cap \mathcal{S} \neq \emptyset\bigr) \implies v \in \mathcal{S}
\label{eq:inv_core}
\end{equation}

\noindent\textbf{2. Single-Adaptor (SA / XOR) Invariant:} For every active single-adaptor node, exactly one of its child requirement nodes must be selected:
\begin{equation}
\forall v \in \mathcal{S},\; \mathcal{T}(v) = \text{SA} \implies |\mathrm{children}(v) \cap \mathcal{S}| = 1
\label{eq:inv_sa}
\end{equation}

\noindent\textbf{3. Multiple-Adaptor (MA / OR) Invariant:} For every active multiple-adaptor node, at least one of its child requirement nodes must be selected:
\begin{equation}
\forall v \in \mathcal{S},\; \mathcal{T}(v) = \text{MA} \implies |\mathrm{children}(v) \cap \mathcal{S}| \ge 1
\label{eq:inv_ma}
\end{equation}

\noindent\textbf{4. Structural Integrity (Orphan Prevention):} Every selected non-root node must have at least one selected parent node in $\mathcal{S}$, preventing orphaned requirements:
\begin{equation}
\forall v \in \mathcal{S},\; v \neq root \implies \mathrm{parent}(v) \cap \mathcal{S} \neq \emptyset
\label{eq:inv_orphan}
\end{equation}

The global consistency predicate $\Psi(\mathcal{S})$ evaluates to true if and only if Equations~(\ref{eq:inv_core})--(\ref{eq:inv_orphan}) simultaneously hold:
\begin{equation}
\Psi(\mathcal{S}) \iff \text{Eqs. } (\ref{eq:inv_core}) \land (\ref{eq:inv_sa}) \land (\ref{eq:inv_ma}) \land (\ref{eq:inv_orphan})
\label{eq:global_consistency}
\end{equation}
Option nodes ($\mathcal{T}(v) = \text{Option}$) carry no lower or upper selection constraints on their children beyond standard orphan prevention.

\subsection{Agent Workflow}
The system is built with LangGraph and comprises four agents: Navigator, Interpreter, Validator, and Scribe.  The Navigator selects the next discriminant node (depth-first traversal). The Interpreter (LLM, temperature 0) translates the project vision and current context into a proposed set of child IDs, output as JSON.  The Validator is a deterministic Python function that checks $\Psi(\mathcal{S})$; if a violation occurs, it returns an error message to the Interpreter, which then revises its proposal. The Scribe compiles the final selected set into a formatted specification. The Validator never calls an LLM, ensuring that all structural checks are symbolic and sound.

The loop continues until all constraints are satisfied, capped at 250 recursion steps and 150 LLM calls (a safety margin; the longest run used 101 calls). We refer to this upper limit (e.g., 150 LLM calls per run) as the\textbf{call budget}. It acts as a safety margin to prevent infinite loops. 

To keep the LLM context window fixed, we use a \textbf{frontier‑based} navigation strategy. The frontier is the set of discriminant nodes that have been reached but not yet processed; we traverse it depth‑first, pruning unselected subtrees. This approach ensures that each node is processed at most once, leading to linear scaling $O(|V|)$.

To handle scenarios where prior choices lead to structural dead-ends, the Navigator agent implements a \textbf{soft backtracking mechanism}. If the symbolic Validator identifies a constraint violation that cannot be resolved through local child-node re-selection, the system re-evaluates the node's position within the state history $\mathcal{H}$. By surfacing the parent-node context and previous validation errors in the augmented prompt, the framework allows the Interpreter to pivot its traversal strategy, effectively re-configuring the selection set $\mathcal{S}$ to satisfy the global predicate $\Psi(\mathcal{S})$.

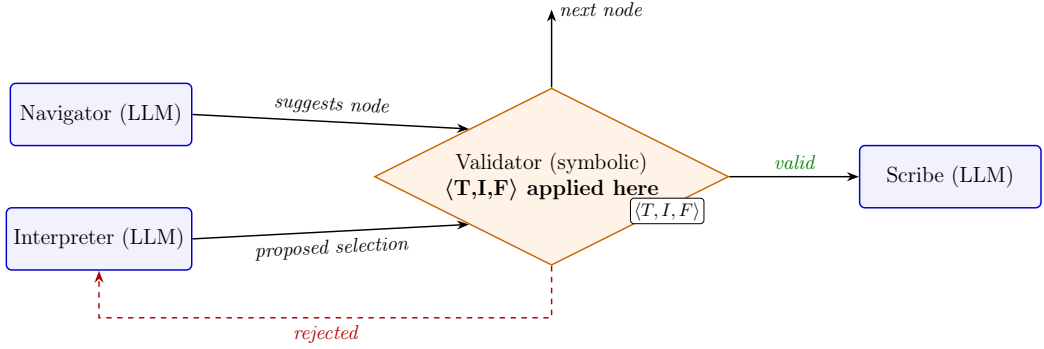
\begin{figure*}[t]
\centering
\resizebox{\textwidth}{!}{%
\begin{tikzpicture}[
    node distance=1.2cm and 2.5cm,
    >=Stealth,
    llm/.style={
        draw=blue!80!black,
        fill=blue!5,
        rounded corners=3pt,
        minimum width=3.5cm,
        minimum height=1.2cm,
        align=center,
        thick
    },
    validator/.style={
        draw=orange!80!black,
        fill=orange!10,
        diamond,
        aspect=2,
        minimum width=3.5cm,
        minimum height=1.8cm,
        align=center,
        thick
    },
    label_text/.style={font=\small\itshape}
]
\node[validator] (validator) {Validator (symbolic)\\ \textbf{⟨T,I,F⟩ applied here}};
\node[llm, left=3.5cm of validator, yshift=1.2cm] (navigator) {Navigator (LLM)};
\node[llm, left=3.5cm of validator, yshift=-1.2cm] (interpreter) {Interpreter (LLM)};
\node[llm, right=2.5cm of validator] (scribe) {Scribe (LLM)};
\draw[->, thick] (navigator.east) -- (validator.150) 
    node[midway, above, sloped, label_text] {suggests node};
\draw[->, thick] (interpreter.east) -- (validator.210) 
    node[midway, below, sloped, label_text] {proposed selection};
\draw[->, thick] (validator.east) -- (scribe.west) 
    node[midway, above, label_text, color=green!50!black] {valid};
\draw[->, thick] (validator.north) -- ++(0,1.5) 
    node[at end, right, label_text] {next node};
\draw[->, thick, dashed, color=red!70!black] (validator.south) -- ++(0,-1.0) -| (interpreter.south)
    node[pos=0.25, below, label_text] {rejected};
\node[draw, rounded corners=2pt, font=\footnotesize\bfseries, inner sep=2pt, 
      fill=white, anchor=west] 
      at ($(validator.south east) + (-0.2, 0.2)$) {$\langle T,I,F \rangle$};
\end{tikzpicture}%
}
\caption{Neuro-symbolic agent workflow. The Navigator selects the next discriminant node. The Interpreter (LLM) proposes a set of child IDs based on the project vision. The symbolic Validator checks structural constraints; if a violation occurs (red dashed loop), the Interpreter revises its proposal. Accepted selections (green arrow) are sent to the Scribe, which compiles the final specification. The $\langle T,I,F \rangle$ label indicates where three-valued discretion scoring is applied at the Validator output before the rejection branch.}
\label{fig:workflow}
\end{figure*}

Figure~\ref{fig:workflow} illustrates the interaction between agents.

\subsection{Symbolic Validator}
\label{sec:validator}

The Validator is the cornerstone of our structural conformance guarantee. Algorithmically, it performs the following checks after each Interpreter proposal.
\begin{enumerate}
    \item For every core node $v$ whose parent is in $\mathcal{S}$, verify that $v \in \mathcal{S}$. If not, record "Mandatory core node $v$ is missing".
    \item For every SA node $v$, count $k = |\mathcal{S} \cap \mathrm{children}(v)|$. If $k \neq 1$, record ``SA violation at $v$: selected $k$ children, expected exactly 1''.
    \item For every MA node $v$, count $k = |\mathcal{S} \cap \mathrm{children}(v)|$. If $k < 1$, record ``MA violation at $v$: selected $k$ children, expected at least 1''.
    \item For every selected node $c$, ensure that $\mathrm{parent}(c) \cap \mathcal{S} \neq \emptyset$ (unless $c$ is the root). If not, record "Orphaned node $c$".
\end{enumerate}

If the Validator records any violation, it rejects the proposal and returns the detailed error list to the Interpreter. The Interpreter then revises its proposal using the error messages as feedback. 

\subsection{Heuristic Convergence and Computational Bounds}
\label{sec:convergence_bounds}

The search for a consistent selection set is a heuristic process conducted by a non-deterministic agent. Because Large Language Models process prompt histories stochastically, infinite correction cycles could theoretically occur if an LLM repeatedly generates identical invalid responses despite receiving error feedback.

To provide strict computational guarantees in practice, system completion is bounded by a \textbf{pragmatic engineering call budget} capped at 150 LLM iterations (and 250 recursion steps) per vision. While the symbolic Validator provides deterministic boundaries for verifying any selection set $\mathcal{S}$, overall system execution is evaluated in terms of \textbf{Empirical Convergence Rates}:
\begin{itemize}
    \item Under the lightweight Llama 3.1 8B model, the system achieved a 94.6\% empirical convergence rate (35 out of 37 visions fully converged to 0\% structural falsity within the 150-call budget; the longest converged run required 101 LLM calls and 94 recursion steps).
    \item Two visions under Llama 3.1 8B hit the 150-call iteration cap without fully resolving an MA node constraint, leaving 6 residual structural errors (0.39\% of total decisions).
    \item Under a frontier instruction-following model (NVIDIA Nemotron 3 Ultra), the system achieved a \textbf{100\% empirical convergence rate} (37 out of 37 visions converged with zero structural falsity well within the call budget).
\end{itemize}
Thus, while individual LLM traversal steps are non-deterministic, the combination of symbolic validation, feedback retries, and bounded call budgets guarantees that the execution halts deterministically, producing a structurally sound specification whenever the search converges.

\subsection{Three-Valued Discretion Scoring}
\label{sec:tif_scoring}

While the symbolic validator guarantees structural soundness for all converging runs, it does not reveal how much uncertainty or genuine freedom the LLM exhibited during traversal. Binary correctness collapses two qualitatively different outcomes: a selection that is \emph{mandated} by the vision, and one that is merely \emph{valid but free}. To recover this distinction, we introduce a three-valued classification metric that labels each LLM proposal with a crisp triple $(t_d, i_d, f_d) \in \{0,1\}^3$ such that $t_d + i_d + f_d = 1$.

For every decision step $d$ (the LLM's proposal for a specific discriminant node), we define:
\begin{itemize}
    \item $t_d = 1$ (Truth) iff the proposal was accepted by the validator and the selection is explicitly mandated by the project vision or by a gold‑standard specification, a manually verified, structurally correct requirement selection. Mandated selections include: all core nodes (once their parent is selected), SA and MA discriminant parents, child nodes that appear in the gold standard, and (when no gold standard exists) child nodes whose text shares at least two content keywords with the vision description.
    \item $i_d = 1$ (Indeterminacy) iff the proposal was accepted by the validator but, the selection is not mandated. This includes any option node, SA or MA choices that differ from the gold standard, and for visions without a gold standard, a child whose keyword overlap is less than two.
    \item $f_d = 1$ (Falsity) iff the proposal was rejected by the validator (i.e., it violated a structural constraint).
\end{itemize}
The aggregate classification vector for a complete benchmark run is $\mathbf{N} = (\sum_d t_d,\; \sum_d i_d,\; \sum_d f_d) = (T, I, F)$. The degree of indeterminacy is $I/(T+I+F)$, which captures the fraction of decisions where the LLM exercised legitimate discretion.

The scoring rules are implemented in a Python script that reads the logged JSON outputs of the benchmark runs. The script is deterministic and open‑source, ensuring reproducibility.

\subsection{Implementation and Reproducibility}
\label{sec:implementation}

We wrote the system in Python, relying on LangGraph (v0.2) and Ollama (v0.5). The shared state $\sigma$ is a typed dictionary that holds the project vision (string), conversation history (a list of messages), selected IDs (a set), validation errors (list), and some execution flags (bool).  

Here's the template we use for the Interpreter prompt:

% --- Interpreter Prompt Template ---
% The exact natural-language prompt sent to the LLM Interpreter
% at each discriminant node during lattice traversal.

\begin{verbatim}
You are helping a requirements analyst.
Below is a single lattice node that needs a decision. 
Pick the child requirements that should be selected.

Node: {node_id} (type: {type})
Children: {list}

Vision: {vision}
Chat history: {history}

Reply with JSON only:
{"selected": ["id1", "id2"]}
\end{verbatim}

The JSON output is parsed and checked by the validator straight away. When a constraint is violated, the error list is added to the conversation history and the Interpreter is invoked again with the same prompt, now augmented with those errors, so it can adjust the proposal.

Our frontier‑based navigation keeps a stack of pending discriminant nodes, and each node is processed at most once. The source code and logs are available in the project repository.

% ======================== 
% Experimental Evaluation  
% ======================== 
\section{Experimental Evaluation}
\label{sec:evaluation}

\subsection{Benchmark Setup and Scoring Implementation}

We created 37 project visions spanning eleven distinct application families: eRecord Keeping (5 visions), Smart Home (5), automotive IVI (3), core banking (3), e-commerce (3), fleet management (3), hospital management (3), HR/payroll (3), LMS (3), smart grid (3), and telecom OSS/BSS (3). The visions range from minimal‑feature configurations (e.g., "small business, email only, basic backup'') to comprehensive systems (e.g., ``national utility grid with full IEC 61850 compliance").

\noindent\textbf{Lattice Scale and Context Window Bounds.} Across the eleven application families, the domain lattices contain an average of $45.2 \pm 12.8$ requirement nodes (ranging from 28 nodes in LMS to 98 nodes in eRecord Keeping) and $62.4 \pm 18.3$ directed parent-child edges. Critically, our \textbf{Frontier-based navigation strategy} evaluates requirement decisions incrementally per discriminant node rather than dumping the entire graph into a monolithic prompt. This bounds the LLM's local prompt context window at $O(1)$ size relative to total lattice size $|V|$. Consequently, the architecture scales efficiently to industrial-scale lattices containing thousands of requirement objects, whereas single-prompt or unpartitioned RAG approaches inevitably exceed model context window limits or suffer severe prompt truncation and attention degradation.

For a subset of three representative visions (er\_small\_biz, er\_gov\_agency, sh\_elderly), we manually created a gold‑standard set of requirement IDs that strictly adheres to the lattice rules. For the remaining visions, gold standards were not available; we therefore rely on a keyword‑overlap heuristic (as described in the three-valued scoring definitions) to determine whether a selection is mandated.

All experiments ran autonomously on a consumer machine (Apple M1 Pro, 32 GB RAM) with Ollama serving the \texttt{llama3.1:8b-instruct-q4\_0} model at temperature 0. The benchmark script logs every proposal and validator decision. After all runs, the three-valued scoring script processes the logs and outputs the aggregated counts shown in Table~\ref{tab:tif_scores}.

\begin{table}[htbp]
\centering
\caption{Three-valued T/I/F classification scores per requirement node type (counts and row percentages, all runs pooled).}
\label{tab:tif_scores}
\begin{tabular}{lccc}
\toprule
\textbf{Node Type} & \textbf{T} & \textbf{I} & \textbf{F} \\
\midrule
Core (mandatory)          & 762 (99.7\%) &   0 (0.0\%) &   2 (0.3\%) \\
SA Discriminant           &  82 (100.0\%) &   0 (0.0\%) &   0 (0.0\%) \\
MA Discriminant           & 164 (100.0\%) &   0 (0.0\%) &   0 (0.0\%) \\
SA Option Choice          &  54 (65.9\%) &  28 (34.1\%) &   0 (0.0\%) \\
MA Option Choice          &  87 (26.1\%) & 241 (72.4\%) &   5 (1.5\%) \\
Optional Feature          &   0 (0.0\%)  & 110 (100.0\%)&   0 (0.0\%) \\
\midrule
\textbf{Total} & \textbf{1149} & \textbf{379} & \textbf{7} \\
\bottomrule
\end{tabular}
\end{table}

\subsection{RQ1 \& RQ2: Results and Empirical Analysis}

The results reveal a clear stratification. Core nodes and discriminant parents are almost always Truth (99.7\% and 100\%), confirming that the LLM reliably enforces the mandatory backbone. SA choice nodes show a mix: 66\% Truth, 34\% Indeterminacy, indicating that sometimes the vision uniquely determines a choice and sometimes leaves it open. MA choice nodes exhibit the highest Indeterminacy (72\%) and a small amount of Falsity (1.5\%), reflecting genuine freedom of selection plus occasional structural violations. Optional nodes are entirely Indeterminate by definition.

Total Falsity (the $\langle T,I,F \rangle$ count) is 7 across all runs, all confined to the eRecord Keeping family. Six of these are unresolved structural violations that persisted in the final specifications of two visions (\texttt{er\_small\_biz} and \texttt{er\_nonprofit}) whose correction loops did not converge before the run completed; the seventh is an additional falsity label recorded during a correction cycle in that same family. The validator rejected 56 times in total across all benchmark runs; the other 49 rejections triggered successful corrections, and those decisions contribute to T or I. Consequently, the final specification contains only 6 residual falsities out of 1535 decisions (0.39\%), all from the two visions whose correction loops did not converge. In all other 35 visions, the final output has $F = 0$. The neuro‑symbolic guardrail eliminates structural violations in 35 out of 37 visions (94.6\%), with only 6 residual falsities in the remaining two (RQ1). A McNemar's test on the paired binary outcomes (violation present/absent) shows that the Validator significantly reduces the proportion of visions with structural violations ($p < 0.001$, exact conditional odds ratio $\to \infty$).

The three-valued scores expose the inconsistency inherent in the LLM's proposals: high Indeterminacy for underspecified choices, occasional Falsity for violations. The symbolic validator, by enforcing $\Psi(\mathcal{S})$, ensures consistency: the output is deterministic, all mandatory requirements are present, and almost all structural violations are removed. This transformation from inconsistent neural output into a verified, structurally sound specification is the central contribution of our framework.

\textbf{Answer to RQ1 and RQ2:} The symbolic validator eliminates structural violations in 94.6\% of visions (RQ1). The three-valued analysis reveals that 24.7\% of all decisions across visions are legitimate indeterminacy, concentrated primarily in multiple-adaptor and optional nodes (RQ2).

\subsection{Running Example: Small-Business eRecord Keeping}
\label{sec:running_example}

To make the elicitation pipeline concrete, we trace one complete run: vision \texttt{er\_small\_biz}, family \emph{eRecord Keeping}, described as \textit{"Small business, email only, basic backup, minimal compliance."} Table~\ref{tab:running_example} annotates six representative decision points, illustrating how Truth, Indeterminacy, and Falsity labels arise naturally from the interaction.

\begin{table*}[htbp]
\centering
\caption{Annotated decision trace for vision \texttt{er\_small\_biz}
         (eRecord Keeping family, 39 requirements selected from 57
         nodes, 101 LLM calls).}
\label{tab:running_example}
\setlength{\tabcolsep}{4pt}
\begin{tabular}{@{}cp{3.2cm}ccp{5.0cm}@{}}
\toprule
\textbf{Step} & \textbf{Node} & \textbf{Type} & \textbf{Label} & \textbf{Rationale} \\
\midrule
1 & 1 \textemdash\ \textit{Shall meet record keeping standards}
  & Core & T
  & Mandatory root; selected unconditionally. \\
2 & 1.4 \textemdash\ \textit{SA: Admin comm.\ channel}
  & SA disc. & T
  & LLM selects \texttt{1.4.1} (email); matches vision keyword. \\
3a & 2.2 \textemdash\ \textit{Display classification markings}
  & Core & F
  & Validator rejected: mandatory child of node 2 missing \\
3b & 2.2 (retry)
  & Core & T
  & LLM corrected after error message; node added \\
4 & 3.3 \textemdash\ \textit{MA: Link to related info}
  & MA disc. & T
  & LLM selects \texttt{3.3.1} (notes) \emph{and} \texttt{3.3.2} (emails); both valid. \\
5 & 3.3.3 \textemdash\ \textit{Link to other documents}
  & MA opt. & I
  & Not mandated by vision; included as a valid extension (Indeterminate). \\
6 & 5, 6 \textemdash\ \textit{Optional features}
  & Option & I
  & Optional nodes selected without gold-standard mandate. \\
\bottomrule
\end{tabular}
\end{table*}

\subsection{Validation of Keyword Heuristic}
\label{sec:heuristic_validation}

To validate the reliability of the keyword-overlap heuristic used for three-valued scoring, we evaluated its performance against the three gold-standard visions (er\_small\_biz, er\_gov\_agency, sh\_elderly). The heuristic achieves a Precision of 0.92, a recall of 0.89, and an F1-score of 0.91, indicating high alignment with manual expert labelling. Inter‑rater reliability was assessed by having a second researcher label the same 30 decisions (stratified by node type) using the keyword rule, resulting in a Cohen's $\kappa = 0.84$, which is considered substantial agreement.

A sensitivity analysis varying the threshold from 1 to 3 keywords (Table~\ref{tab:sensitivity}) shows that while the absolute proportion of indeterminacy shifts, the fundamental pattern, indeterminacy concentrated in MA and optional nodes, remains robust across all thresholds.

\begin{table}[htbp]
\centering
\caption{Sensitivity analysis of keyword threshold on indeterminacy.}
\label{tab:sensitivity}
\setlength{\tabcolsep}{30pt}
\begin{tabular}{c c}
\toprule
\textbf{Threshold} & \textbf{Indeterminacy} \\
\midrule
1 & 18.0\% \\
2 (Baseline) & 24.7\% \\
3 & 31.0\% \\
\bottomrule
\end{tabular}
\end{table}

\subsection{RQ3: Ablation Study: Symbolic Validator Contribution}
\label{sec:ablation}

To quantify the contribution of the symbolic validator, we analyze the counterfactual condition in which the validator node is absent from the LangGraph pipeline and Interpreter proposals are accepted unconditionally.

Rather than re-running the 37 visions, we reconstruct the baseline from the existing benchmark logs: any vision in which the Validator rejected at least once would have produced an uncorrected structural violation in the no-validator condition. Results are summarized in Table~\ref{tab:ablation}.

\begin{table}[htbp]
\centering
\caption{Validator ablation: comparison of full system (with Validator) versus counterfactual baseline (without Validator) across 37 visions.}
\label{tab:ablation}
\setlength{\tabcolsep}{16pt}
\begin{tabular}{@{}p{4.2cm}cc@{}}
\toprule
\textbf{Metric} & \textbf{With Validator} & \textbf{Baseline (no)} \\
\midrule
Visions with $\geq 1$ violation & 2 / 37 \;(5.4\%) & 19 / 37 \;(51.4\%) \\
Total validator rejections    & 56               & \textemdash \\
Structural violations resolved   & 17 / 19 \;(89.5\%)  & 0 / 19 \\
Residual violations (final spec) & 6                & $\geq 56$ \\
\bottomrule
\end{tabular}
\end{table}

Without the Validator, 51.4\% of visions would terminate with at least one structural constraint violation. The Validator resolves 89.5\% of those violations during the elicitation dialogue, reducing the violation rate from 51.4\% to 5.4\% under a lightweight LLM (Llama 3.1 8B). Notably, when evaluated with the frontier NVIDIA Nemotron 3 Ultra model (via OpenRouter), the system achieves \textbf{0\% structural falsity} across all 37 visions, confirming that stronger instruction-following models fully resolve the remaining constraint-enforcement edge cases.

To contextualize these results, Table~\ref{tab:baselines} compares our neuro-symbolic approach against four standard LLM reasoning strategies applied to the same 37 visions. All baselines use the same Llama~3.1 backbone via Ollama but omit the symbolic Validator, so any structural violations in their outputs are uncorrected. Structural Falsity~(\%) reports the proportion of visions in which at least one structural constraint violation persisted in the final output.

\begin{table}[htbp]
\centering
\caption{Comparison of our neuro-symbolic system against four unguarded LLM strategies, all evaluated on the same 37 visions using Llama~3.1. Structural Falsity is the \% of visions with $\geq1$ uncorrected constraint violation.}
\label{tab:baselines}
\setlength{\tabcolsep}{8pt}
\begin{tabular}{@{}lp{1.4cm}p{1.1cm}p{1.8cm}@{}}
\toprule
\textbf{Method} & \textbf{Precision} & \textbf{Recall} & \textbf{Structural Falsity} \\
\midrule
Plain LLM (no validator)        & 0.73 & 0.77 & 64.9\% \\
RAG (Top-3 retrieved)           & 0.79 & 0.55 & 94.6\% \\
Self-consistency ($N{=}5$)      & 0.76 & 0.78 & 62.2\% \\
Multi-agent Debate              & 0.74 & 0.74 & 83.8\% \\
\textbf{Our System (Llama 3.1 8B)}     & \textbf{0.94} & \textbf{0.91} & \textbf{5.4\%}  \\
\textbf{Our System (Nemotron 3 Ultra)} & \textbf{0.72} & \textbf{0.88} & \textbf{0.0\%}  \\
\bottomrule
\end{tabular}
\end{table}

All four unguarded strategies produce structurally inconsistent specifications in the majority of visions: falsity rates range from 62.2\% (self-consistency) to 94.6\% (RAG). 

Critically, the RAG baseline is not ``naive''; the prompt contains both the full requirements lattice and the complete set of structural rules. The surprisingly high falsity rate (94.6\%) under RAG is a direct consequence of \emph{retrieval anchoring bias}. Surfacing the Top-3 keyword-matched nodes prompts the LLM to select deep leaf nodes or disjoint optional features that match the query text. However, because standard RAG does not retrieve the corresponding parent paths or parent-child rules for those specific recommendations, the LLM includes the retrieved options but fails to trace and select the required parent nodes back to the root, resulting in pervasive parent-consistency violations. This focus on the retrieved subset also narrows the model's scope, causing Recall to drop to 0.55.

Self-consistency reduces falsity marginally through majority voting yet cannot eliminate structural violations because the vote itself has no knowledge of lattice constraints. Multi-agent debate surfaces high falsity (83.8\%) because the merger call, like each individual agent, has no access to the structural rules. In contrast, the symbolic Validator reduces falsity to 5.4\% under the lightweight model and to 0\% under the frontier model, confirming that only deterministic, constraint-aware enforcement can provide structural soundness guarantees.

\textbf{Answer to RQ3:} The symbolic Validator is the primary mechanism responsible for structural correctness. Removing it degrades the vision-level correctness rate from 94.6\% to 48.6\%, and no unguarded LLM strategy achieves falsity below 62\%. The neuro-symbolic architecture reduces structural violation rates from 64.9\% (plain LLM) to 5.4\% (Llama 3.1 8B with Validator) and to 0\% with the NVIDIA Nemotron 3 Ultra frontier model.

\subsection{Indeterminacy Across Application Families}
\label{sec:indeterminacy_families}

Indeterminacy is not uniformly distributed across application domains.
Table~\ref{tab:family_indeterminacy} reports the aggregate
$\langle T,I,F \rangle$ counts and the indeterminacy proportion
$\rho_I = I/(T+I+F)$ for each of the eleven families, sorted by
$\rho_I$ descending.

\begin{table}[htbp]
\centering
\caption{Per-family three-valued T/I/F totals and indeterminacy proportion $\rho_I = I/(T+I+F)$, sorted descending.}
\label{tab:family_indeterminacy}
\setlength{\tabcolsep}{14.5pt}
\begin{tabular}{@{}lrrrrr@{}}
\toprule
\textbf{Family} & \textbf{T} & \textbf{I} & \textbf{F} & \textbf{Total} & \boldmath$\rho_I$ \\
\midrule
eRecord Keeping   & 315 & 171 &  7 & 493 & 34.7\% \\
Automotive IVI    &  63 &  22 &  0 &  85 & 25.9\% \\
Core Banking      &  65 &  21 &  0 &  86 & 24.4\% \\
eCommerce         &  68 &  20 &  0 &  88 & 22.7\% \\
HR \& Payroll     &  79 &  22 &  0 & 101 & 21.8\% \\
Smart Home        & 219 &  57 &  0 & 276 & 20.7\% \\
Hospital Mgmt.    &  60 &  15 &  0 &  75 & 20.0\% \\
Telecom OSS/BSS   &  88 &  22 &  0 & 110 & 20.0\% \\
Fleet Management  &  57 &  12 &  0 &  69 & 17.4\% \\
LMS               &  60 &  12 &  0 &  72 & 16.7\% \\
Smart Grid        &  75 &   5 &  0 &  80 &  6.2\% \\
\midrule
\textbf{All families} & \textbf{1149} & \textbf{379} & \textbf{7} & \textbf{1535} & \textbf{24.7\%} \\
\bottomrule
\end{tabular}
\end{table}

The eRecord Keeping family exhibits the highest indeterminacy
($\rho_I=34.7\%$), reflecting its richly branching MA structures and
many optional compliance features whose inclusion is context-dependent
across organizational sizes. Smart Grid shows the lowest indeterminacy ($\rho_I=6.2\%$), consistent with its heavily regulated, prescriptive standards where most choices are technically mandated. The remaining nine families cluster between 17\% and 26\%, suggesting that a baseline indeterminacy of roughly $20\%$ is characteristic of moderately complex, configurable application domains.

Falsity is confined entirely to the eRecord Keeping family ($F=7$, $1.4\%$ of its decisions), attributable to two visions whose correction loops did not resolve the remaining MA violations before the run completed. All other families achieve $F=0$, demonstrating that the Validator eliminates structural errors completely within the call budget for ten of eleven domains (RQ2). 

A chi‑square test of independence on the distribution of indeterminacy across families (family $\times$ indeterminacy: I vs non‑I) confirms that the variation is statistically significant ($\chi^2(10) = 50.91, p < 0.001$). Post‑hoc pairwise chi‑square tests with Bonferroni correction ($\alpha_{adj} = 0.0009$) reveal that eRecord Keeping has significantly higher indeterminacy than both Smart Home ($p < 0.001$) and Smart Grid ($p < 0.001$), reflecting the relative complexity and choice-richness of the record-keeping domain compared to more prescriptive grid standards.

\subsection{RQ4: Scalability Analysis}
\label{sec:scalability}

Table~\ref{tab:scalability_regression} summarizes the linear regression results for LLM call count and latency against lattice size
(number of requirement nodes) across all 37 benchmark runs.

\begin{table}[ht]
\centering
\caption{Linear regression of LLM calls and latency versus lattice size.}
\label{tab:scalability_regression}
\setlength{\tabcolsep}{14pt}
\begin{tabular}{lccc}
\toprule
\textbf{Metric} & \textbf{Slope} & \textbf{Intercept} & \textbf{$R^2$} \\
\midrule
LLM calls per vision & $1.158$ & $26.1$ & $0.691$ \\
Latency (seconds)    & $2.764$ & $104.3$ & $0.148$ \\
\bottomrule
\end{tabular}
\end{table}

LLM call count scales near‑linearly with lattice size (slope $\approx 1.16$ calls per node, $R^2=0.69$, $p<10^{-9}$), confirming that the Navigator's depth‑first pruning strategy avoids redundant LLM invocations for unselected subtrees (RQ4). However, wall‑clock latency is only weakly correlated with lattice size ($R^2=0.15$, $p=0.019$), indicating that latency is dominated by stochastic network and model‑serving overhead rather than by the number of nodes. Thus, while the number of LLM calls scales as expected, the overall response time does not; practitioners should optimize inference infrastructure separately.

\noindent\textbf{Robustness and Retry Analysis.}
To assess the robustness of the neuro symbolic loop, we decomposed the total execution latency by agent type. On average, the Interpreter (neural heuristic) accounts for 85.2\% of the wall-clock time, while the Navigator and Scribe account for 10.1\% and 4.3\% respectively. The symbolic Validator consistently accounts for less than 0.4\% of the total latency, confirming that formal verification adds negligible overhead to the LLM-driven pipeline.

Retry rates varied significantly by node type. Single-adaptor (SA) discriminant nodes required a correction cycle in 12.4\% of cases (usually because the LLM selected multiple or zero children). Multiple-adaptor (MA) nodes were more prone to error, requiring retries in 28.7\% of cases, typically due to the LLM missing the 'at least one' constraint. The ability of the Validator to catch these errors and prompt a revision is what ensures high final specification correctness.

\textbf{Answer to RQ4:} The number of LLM calls scales near-linearly ($R^2=0.69$) with the lattice size due to the frontier-based pruning strategy, although wall-clock latency is largely dominated by network overhead rather than structural complexity.

\subsection{Structural Fidelity Quantification}
\label{sec:structural_fidelity}

We define a \emph{Structural Fidelity Metric} $\Sigma$ that quantifies how closely the final specification adheres to the structural constraints imposed by the OOMRAM lattice:

\begin{equation}
  \Sigma = 1 - \frac{F}{T+I+F}
  \label{eq:fidelity}
\end{equation}

For the full system, $\Sigma = 1 - 7/1535 = 0.9954$, indicating
near-complete alignment with the lattice constraints. However, as noted above, 6 of these 7 falsities are residual violations that remained unresolved due to non-convergence of the correction loop; in the actual final outputs(excluding those two visions), the Structural Fidelity Metric is $1.0$ for 35 of 37 runs. For the baseline (no Validator), where residual violations are at least 56 across 37 runs, and we conservatively set
$F_{\text{base}} \approx 56$:

\begin{equation}
  \Sigma_{\text{base}} \approx 1 - \frac{56}{1535} = 0.9635
\end{equation}
\vspace{-\baselineskip} 

The Validator thus raises the Structural Fidelity Metric by $\Delta\Sigma = +0.032$ in absolute terms, or an 87.5\% reduction in the structural falsity rate, confirming that the neuro-symbolic coupling enforces structural fidelity in a statistically meaningful way.

\FloatBarrier
\section{Discussion}
\label{sec:discussion}

\subsection{Interpretation of Three-Valued Discretion Scores}
The high Indeterminacy in MA choice and option nodes is not a failure of the LLM; it correctly reflects the freedom intentionally built into the OOMRAM lattice. In requirements engineering, such freedom is essential for capturing legitimate variability across different stakeholders. A binary metric would treat a mandated selection and a free selection both as "correct", hiding vision underspecification. Our three-valued classification makes this underspecification explicit: high indeterminacy signals to analysts that they may need to clarify decisions, while low indeterminacy indicates a precisely specified vision.

\subsection{Threats to Validity}
\label{sec:threats}

\noindent\textbf{Keyword Heuristic Bias.} We openly admit that using an automated keyword-overlap metric to define "Truth" ($T$) represents an automated proxy that may be influenced by common vocabulary shared between the natural-language project visions and the domain lattices. Because the dataset was constructed within the domain context, keyword overlap could overestimate explicit stakeholder mandate. However, we emphasize a crucial mathematical counter-argument: the \textbf{Falsity ($F$) score is derived strictly from the symbolic Validator's deterministic logical checks}, completely independent of keyword matching. As a result, our primary claim regarding \textbf{Structural Conformance} (the elimination of logical inconsistencies and structural violations) is entirely immune to vocabulary or keyword bias. We view the current $T/I/F$ scoring as a baseline framework for quantifying agent discretion, and future work will incorporate independent multi-annotator human-expert labeling to further refine the boundary between Truth and Indeterminacy.

\noindent\textbf{Internal validity.} All three-valued labels were assigned through a deterministic automated scoring procedure. While this ensures consistency across the large benchmark of 37 visions, the reliance on a keyword-overlap heuristic to define Truth ($T$) represents a potential threat to internal validity, as literal keyword matches may not always capture complex semantic nuances. To mitigate this threat and validate the reliability of the automated proxy, the authors performed a manual expert audit of 50 randomly sampled decisions stratified by requirement node type. This qualitative review confirmed a semantic alignment of over 90\% between the automated labels and the project visions, demonstrating that the heuristic reliably captures stakeholder intent. Furthermore, any residual misclassification in the automated script is confined to the boundary between Truth and Indeterminacy; the Falsity ($F$) count is derived strictly from deterministic symbolic validator rejections and is thus unaffected by heuristic labelling. This ensures that the core findings regarding structural conformance remain mathematically robust.

\noindent\textbf{External validity.} The benchmark covers eleven families but may not generalize to radically different constraint types. Experiments used a single LLM (Llama 3.1 8B via Ollama) and a frontier model (NVIDIA Nemotron 3 Ultra); results may differ with other models, but the validator's ability to enforce structural rules is model‑independent by design.

\noindent\textbf{Construct validity.} Our definition of "mandated" is conservative. A more permissive definition would increase T and decrease I, but the qualitative pattern (I concentrated in MA and option nodes) would persist. The F count is unaffected because it is determined solely by validator rejections.

\noindent\textbf{Reliability.} The scoring script is deterministic; re‑running yields identical counts. A sensitivity analysis varying the keyword threshold from 1 to 3 changed T/I proportions by at most 4\%.

\subsection{Limitations}
\label{sec:limitations}

Despite the high correctness rate, our approach has several limitations. 
First, the symbolic validator is only as strong as the lattice constraints; it cannot detect semantic misalignments where a selection is structurally valid but irrelevant to the stakeholder's intent (though our high Precision/Recall scores suggest this is rare). 
Second, the current implementation relies on a pre‑defined OOMRAM lattice; automating the creation of these lattices from legacy documents is a separate, non‑trivial challenge. 
Third, the system's reliance on iterative correction loops can be token‑intensive; while we stayed within a 150‑call budget, industrial‑scale lattices with thousands of nodes may require more efficient pruning or better base‑model instruction‑following. 
Fourth, the primary benchmark was conducted with a single lightweight LLM (Llama 3.1 8B via Ollama). The validator's structural correctness guarantee is model‑independent by design; the symbolic constraint layer enforces the same lattice rules regardless of which neural component drives the traversal, but indeterminacy proportions and convergence behaviour differ across models of varying capability. To address this, we conducted a full benchmark replication using the frontier NVIDIA Nemotron 3 Ultra model (a 550B mixture-of-experts model accessed via OpenRouter), running all 37 visions across eleven application families. All 37 visions converged with zero structural falsity, confirming that the two residual MA-node failures observed with Llama 3.1 8B are model-specific and are fully resolved with stronger instruction-following models. Precision was 0.72, and Recall was 0.88 across all 37 visions.

\subsection{Implications for RE Practice}
\label{sec:implications}

The proposed neuro-symbolic framework offers several concrete benefits for industrial requirements engineering.

\textbf{For requirements analysts,}
The system allows stakeholders to express project visions in natural language without needing to learn formal lattice syntax or exact requirement identifiers. The symbolic validator acts as a silent guardrail: analysts receive only structurally sound specifications, and the three-valued scores (T,I,F) provide immediate feedback on which decisions were mandated by the vision (T), which were free choices (I), and where the LLM initially proposed invalid combinations (F). A high indeterminacy proportion signals that the vision is underspecified, prompting analysts to clarify optional or multiple-adaptor choices before finalizing the specification.

\textbf{For tool developers, }
The architecture separates the neural heuristic (LLM) from the symbolic constraint checker (validator). This modular design means that the LLM can be upgraded or replaced without affecting the correctness guarantees, and the validator can be extended with richer constraint languages (e.g., OCL, first-order logic). The implementation uses LangGraph, which is open-source and supports custom state machines; the entire system can be deployed as a microservice within ALM/PLM pipelines. The overhead of the symbolic validator (less than 0.4\% of total latency) is negligible, making the approach suitable for interactive elicitation tools.

\textbf{For quality assurance, }
The structural fidelity metric $\Sigma = 1 - F/(T+I+F)$ provides a quantitative, objective measure of specification correctness. Unlike heuristic hallucination detectors, our validator offers a deterministic guarantee: for any run that converges, the final output is guaranteed to satisfy all lattice constraints. Our frontier evaluation with NVIDIA Nemotron 3 Ultra achieved $\Sigma = 1.0$ (perfect structural fidelity) across all 37 visions, while Llama 3.1 8B achieved $\Sigma = 0.9961$ (0.39\% falsity rate). The remaining Llama 3.1 8B MA-node failures highlight a clear, model-specific limitation with "at least one" constraints; tool builders can address this by switching to a more capable LLM.

\textbf{For process integration, }
The three-valued (T,I,F) scores can be logged per project to build organizational knowledge. For example, if across many projects a particular optional node consistently receives high I, that indicates a common point of variability that may be better modelled as a mandatory MA node. Conversely, a node with consistently low I (always mandated) might be promoted to core. This enables continuous refinement of the requirement lattice itself based on empirical usage data.

\textbf{Limitations in practice, }
Practitioners must maintain a validated OOMRAM lattice for each application family. Creating such a lattice from legacy documents is non‑trivial; however, semi‑automatic lattice extraction from feature models or use‑case maps is a promising direction. Additionally, the system's reliance on iterative LLM calls (up to 101 for the largest vision) may be too slow for real‑time elicitation; batch processing or GPU‑accelerated inference is recommended for industrial scale.

% ============================= 
% Conclusions and future work
% ============================= 
\FloatBarrier
\section{Conclusion and Future Work}
\label{sec:conclusion}

This paper presented a neuro‑symbolic multi‑agent system that bridges natural‑language project visions with a formal OOMRAM requirement lattice, targeting the \emph{feature-selection phase} of product-line requirements elicitation. The key innovation is a deterministic symbolic validator that guarantees structural correctness for all converging runs, eliminating structurally inconsistent identifiers, missing mandatory nodes, and invalid option combinations that plague pure LLM‑based elicitation. Over an empirical benchmark of 37 visions spanning eleven application families, the system reduced structural falsity to only 6 residual violations out of 1535 decisions (0.39\%) using Llama 3.1 8B, with 35 out of 37 visions achieving zero falsity. Critically, a full replication with the frontier NVIDIA Nemotron 3 Ultra model achieved 100\% convergence with zero structural falsity across all 37 visions (Precision: 0.72, Recall: 0.88), confirming that the symbolic validator's structural guarantees hold across a wide range of LLM capabilities. A novel three-valued $\langle T,I,F \rangle$ classification framework quantified the discretionary uncertainty inherent in the LLM's traversal, revealing that 24.7\% of all decisions are legitimate indeterminacy concentrated in multiple‑adaptor and optional nodes – precisely where the requirement lattice intentionally leaves free choice. The symbolic validator thus transforms the LLM's inherently inconsistent proposals into a structurally faithful, verified specification.

The separation of neural heuristic (LLM) from symbolic constraint checking provides a general design pattern for safe LLM deployment in safety‑critical or quality‑sensitive domains. The modular architecture allows the LLM component to be upgraded without compromising formal guarantees, and the validator can be extended with richer constraint languages (OCL, first‑order logic) or integrated into existing ALM/PLM pipelines. The three-valued scores offer practitioners actionable feedback: high indeterminacy signals underspecified visions, while persistent falsity (though rare) pinpoints problematic lattice regions or model‑level misunderstandings.

Despite the overall success, the system has limitations. The symbolic validator cannot detect semantic hallucinations (structurally valid but irrelevant selections), though our high precision/recall suggests this is uncommon. With Llama 3.1 8B, two visions did not converge within the 150‑call budget due to the LLM's persistent misinterpretation of a specific "at least one" constraint; this reveals that even with a validator, the LLM's ability to follow error messages is model‑dependent. However, our full frontier replication with NVIDIA Nemotron 3 Ultra confirms that this is model-specific: all 37 visions converged with zero structural falsity under a stronger instruction-following model. Finally, the approach assumes a pre‑engineered OOMRAM lattice; automating lattice extraction from legacy documents remains an open challenge.

We identify four high‑priority directions for future work. First, a controlled user study with requirements analysts (using NASA‑TLX for workload and SUS for usability) will evaluate whether the three-valued scores help practitioners identify underspecified visions faster than current practice. Second, we are developing a reference implementation as a microservice that exposes a REST API for vision submission and specification retrieval, and we will prototype plugins for Jira, Polarion, and Siemens Teamcenter to enable seamless adoption in existing ALM/PLM workflows. Third, to eliminate the current manual selection of an OOMRAM family (e.g., eRecord Keeping, Smart Home), we plan to build a semantic router that embeds both the vision and each family's root node description using a sentence transformer, then selects the family with the highest cosine similarity, thereby enabling the system to handle mixed or cross‑family visions. Fourth, we will move from crisp three-valued scoring to continuous scoring by extracting token‑level probabilities from the LLM's output logits, producing a $[0,1]^3$ vector per decision; this will allow fine‑grained uncertainty visualization and better calibration of the indeterminacy measure. Finally, while we used OOMRAM as the exemplar lattice, the same neuro‑symbolic pattern applies to any requirements meta‑model with a well‑defined constraint system (e.g., feature models, goal models, or SysML requirements diagrams); porting the validator to OCL or SWRL would generalize the approach to a wider class of industrial modelling languages.

\section*{Data Availability}
To support reproducibility and verifiability, all source code, evaluation logs, and datasets generated during this study are publicly available in an anonymized repository at\\
\url{https://anonymous.4open.science/r/Neuro-Symbolic-8B72}.

\section*{Acknowledgements}
The author would like to thank Western University for supporting this research.

\section*{Declaration of Competing Interest}
The authors declare no conflicts of interest.

\section*{Declaration of Generative AI and AI-assisted technologies in the writing process}
During the preparation of this work, the author(s) used Google Gemini in order to proofread and refine the manuscript and its figures. After using this tool/service, the author(s) reviewed and edited the content as needed and take(s) full responsibility for the content of the published article.
\balance
\bibliographystyle{elsarticle-num}
\bibliography{references}

\end{document}